\documentclass[twocolumn]{aastex63}
\usepackage{graphicx}
\usepackage{amsmath}
\usepackage{apjfonts,natbib}
\usepackage{appendix}

\newcommand{\gsim}{\lower.5ex\hbox{$\; \buildrel > \over \sim \;$}}
\newcommand{\lsim}{\lower.5ex\hbox{$\; \buildrel < \over \sim \;$}}

\newcommand{\ciii}{\hbox{C\,{\sc iii}}}
\newcommand{\civ}{\hbox{C\,{\sc iv}}}
\newcommand{\cv}{\hbox{C\,{\sc v}}}

\newcommand{\kms}{\ifmmode {\rm km\,s}^{-1} \else km\,s$^{-1}$ \fi}
\newcommand{\cc}{\hbox{cm$^{-3}$}}

\newcommand{\ergcms}{\ifmmode {\rm ergs\,cm}^{-2}\,{\rm s}^{-1} \else ergs\,cm$^{-2}$\,s$^{-1}$\fi}
\newcommand{\ergcmsA}{\ifmmode{\rm ergs}\, {\rm cm}^{-2}\,{\rm s}^{-1}\,{\rm\AA}^{-1} \else ergs\, cm$^{-2}$\, s$^{-1}$\, \AA$^{-1}$\fi}
\newcommand{\ergcmsHz}{\ifmmode{\rm ergs\,cm}^{-2}\,{\rm s}^{-1}\,{\rm Hz}^{-1} \else ergs\,cm$^{-2}$\,s$^{-1}$\,Hz$^{-1}$\fi}
\newcommand{\phcms}{\ifmmode {\rm ph\,cm}^{-2}\,{\rm s}^{-1} \else ,ph\,cm$^{-2}$\,s$^{-1}$\fi}
\newcommand{\phcmsA}{\ifmmode {\rm ph\,cm}^{-2}\,{\rm s}^{-1}\,{\rm\AA}^{-1} \else ph\,cm$^{-2}$\,s$^{-1}$\,\AA$^{-1}$\fi}
\newcommand{\Luv}{\ifmmode L_{1450} \else $L_{1450}$\fi}
\newcommand{\Lop}{\ifmmode L_{5100} \else $L_{5100}$\fi}
\newcommand{\Lthree}{\ifmmode L_{3000} \else $L_{3000}$\fi}
\newcommand{\lledd}{\ifmmode L/L_{\rm Edd} \else $L/L_{\rm Edd}$\fi}
\newcommand{\ledd}{\ifmmode L_{\rm Edd} \else $L_{\rm Edd}$\fi}
\newcommand{\lamLlam}{\ifmmode \lambda L_{\lambda} \else $\lambda L_{\lambda}$\fi}
\newcommand{\lbol} {\ifmmode L_{\rm bol} \else $L_{\rm bol}$\fi}
\newcommand{\llbol}{\ifmmode \log\left(\lbol/\ergs\right) \else $\log\left(\lbol/\ergs\right)$\fi}
\def\dif{\mathop{}\hphantom{\mskip-\thinmuskip}\mathrm{d}}%
\let\daccent\d
\gdef\d{\ifmmode\dif\else\expandafter\daccent\fi}

\begin{document}

\title{Discovery of the Hybrid Response of Photoionized Gases}
\shortauthors{He, Wang \& Ferland}
\shorttitle{The Hybrid Response}

\correspondingauthor{Zhicheng He}
	\email{zcho@ustc.edu.cn}
\author[0000-0003-3667-1060]{Zhicheng He}
\affiliation{School of Astronomy and Space Science, University of Science and Technology of China, Hefei, Anhui 230026, China
}%
\correspondingauthor{Tinggui Wang}
\author{Tinggui Wang}%
\email{twang@ustc.edu.cn}
\affiliation{School of Astronomy and Space Science, University of Science and Technology of China, Hefei, Anhui 230026, China
}%
\correspondingauthor{Gary J. Ferland}
\author{Gary J. Ferland}%
\email{gary@g.uky.edu}
\affiliation{Department of Physics and Astronomy, The University of Kentucky, Lexington, KY 40506, USA
}%

\date{\today}

\begin{abstract}
Photoionized gases are prevalent throughout the universe. In such gases, the ion concentration typically exhibits two response modes to radiation: a positive response in the low-ionization state and a negative response in the high-ionization state. Here, we report the discovery of a widespread misalignment at the boundary between the above two response modes, and identify a third mode—the hybrid response—through time-dependent photoionization simulations. This phenomenon arises from the asynchrony among the ionization rate, recombination rate, and ion column density. Among these, only the ionization rate can respond instantaneously to changes in radiation. Consequently, the initial rate of change in the column density of \( N_i \) ion is given by \( -N_i I_i + N_{i-1} I_{i-1} \). However, this quantity is typically nonzero at the peak of \( N_i \), leading to a misalignment between the boundaries of positive and negative responses. Such hybrid effects introduce additional complexity in the interpretation of gas properties, highlighting the need for further investigation.
\end{abstract}



\section{\label{sec:intro}Introduction}

Ionized gas (also known as a plasma) in regions such as the interstellar medium (ISM) and intergalactic medium (IGM), is prevalent in the universe, especially  after the cosmic reionization epoch with $z<6$ 
\citep{barkana2001,kriss2001,zheng2004,fan2006,tilvi2020,yung2020}. 
Atoms undergo ionization primarily through two mechanisms: collisional ionization and photoionization. Collisional ionization occurs when the thermal energy $\mathrm{k}T$ ($\mathrm{k}$ is the Boltzmann constant and $T$ is the gas temperature) of particles is comparable to the ionization energy of the ions. 
In general, in high-density gases, the ion concentration dominated by collisional ionization can be described by the Saha equation\cite{saha1920,saha1921} under local thermodynamic equilibrium (LTE), except for some optically thin low-density cases, such as the stellar corona. Photoionization occurs when the energy of incident photons exceeds the ionization energy of the ions. Ionizing sources in the universe, including stars and quasars (a class of high luminosity active galactic nuclei powered by a supermassive black hole, or abbreviated as SMBH), make photoionization processes widespread. For example, the famous $R-L$ relation \cite{kaspi2005,bentz2006,bentz2009,du2019} between the size of the broad-line region (BLR) and the SMBH-accretion disk luminosity, i.e., $R\propto L^{0.5}$ is well explained by the photoionization model \cite{korista2000,baskin2014,guo2020,wu2024}. This $R-L$ relation underlies the SMBH mass estimation with single-epoch spectroscopy, using luminosity as a proxy for the BLR size \cite{shen2011}, and can also be used to measure cosmological distance \cite{watson2011}. Studies of absorption lines in quasar spectra have shown that ions such as $\rm N^{4+}$, $\rm C^{3+}$, and $\rm Si^{3+}$ in the gas surrounding quasars are primarily governed by photoionization
(e.g., \cite{wang2015,he2017,lu2018,hemler2019,vivek2019,zhao2021}).
quasar radiation typically exhibits random variability \citep{kelly2009}, providing an excellent opportunity to infer the ionization parameters, density, temperature, and other properties of the gas by analyzing how the ionization states of the gas respond to changes in radiation. However, the response of gas ionization states to radiation is a complex, non-equilibrium evolution process that remains not fully understood. Here we will introduce a puzzling problem we've encountered, leading to the discovery of an intriguing new phenomenon.

\begin{figure*}
\centering
\includegraphics[width=0.9\textwidth]{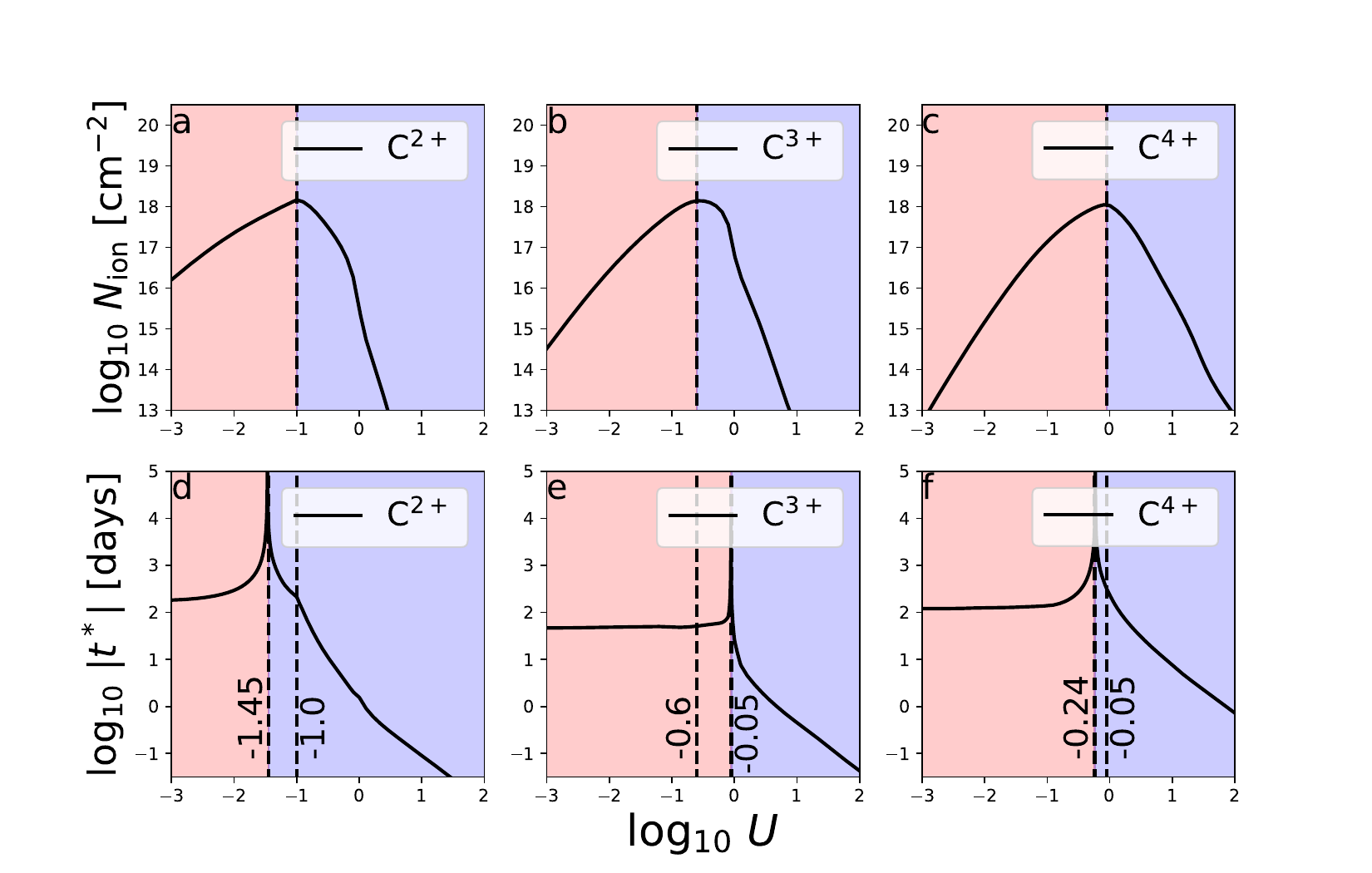}
\caption{\textbf{The Phenomenon of Positive and Negative Response Misalignment for Carbon Ions in Photoionized Gases.} The top panels show the ion column density as a function of ionization parameters, while the bottom panels display the characteristic timescale. Dashed lines indicate the boundaries of positive and negative responses, determined by the peak column density and characteristic timescale, respectively, highlighting their misalignment. The light red region represents the positive response, while the light blue region indicates the negative response. Detailed parameter settings for the photoionization simulation are provided in Section \ref{sec:parameter}.
\label{fig1}}
\end{figure*}

In photoionized gases, the population density of a specific ion initially increases as the ionization parameter rises. Eventually, it reaches a peak before starting to decrease. The ionization parameter is defined as $U = Q_{\rm H}/[4\pi r^2n_{\rm H}c]$, where $Q_{\rm H}$ represents the emission rate of hydrogen-ionizing photons, $r$ is the distance to the gas from the source, $c$ is the speed of light, and $n_{\rm H}$ is the hydrogen number density.
As shown in panel a, b and c of Fig. \ref{fig1}, we used CLOUDY \citep{ferland2017} simulations (see details in the next section) to demonstrate that the column density of ions $\rm C^{2+}$ (ionization energy 47.9 eV), $\rm C^{3+}$ (64.5 eV), and $\rm C^{4+}$ (392.1 eV) first increases with ionization parameters, reaches a peak, and then decreases as ionization parameters continue to increase.
Naturally, the ionization parameter corresponding to the peak increases as the ionization energy increases. 
We will refer to the left side as the low-ionization state (represented by the light red region in Fig. \ref{fig1}) and the right side as the high-ionization state (represented by the light blue region). In the low-ionization state, there is a positive correlation, termed a positive response, between the ion column density and the ionization parameter. In contrast, in the high-ionized region, this relationship is reversed, showing a negative response. 

The column density of of a given element in ionization stage i is represented by 
\begin{eqnarray}  \label{eq1}
\frac{\d N_i}{\d t}=-N_i(I_i+R_{i-1})+N_{i-1}I_{i-1}+N_{i+1}R_{i},
\end{eqnarray}
where the ionization rate per particle is $I_i$, and the recombination rate per particle from ionization stage i+1 to i is 
$R_i= \alpha_i(T)n_e$. The recombination coefficient $\alpha_i$ depends on the electron temperature $T$ \citep{osterbrock2006}.
In the photoionization equilibrium, these reduce to n equations of the form
\begin{eqnarray}  \label{eq2}
\frac{N_{i+1}}{N_{i}}=\frac{I_{i}}{R_{i}}.
\end{eqnarray}
This implies that the increase of stage i by recombination from stage i+1 must be balanced by the decrease of stage i by ionization to stage i+1. Suppose a gas in photoionization equilibrium experiences a sudden change in the incident ionizing flux such that $I_i(t > 0)=(1+f)I_i(t=0)$, 
where $-1\le f \le +\infty$ (refer to \citealt{arav2012}). The characteristic timescale \citep{krolik1995, nicastro1999, bottorff2000, arav2012, netzer2013} for change in the ionic column density is:
\begin{eqnarray} 
t_i^*=\left[-f \alpha_{i} n_{e}\left(\frac{N_{i+1}}{N_{i}}-\frac{\alpha_{i-1}}{\alpha_{i}}\right)\right]^{-1}.
\label{eq3}
\end{eqnarray}
$t_i^*$ is positive if $\frac{N_{i+1}}{N_{i}}$ is less than $\frac{\alpha_{i-1}}{\alpha_{i}}$, negative if greater, and approaches infinity if equal. These conditions lead to a positive, negative, or no response in the ionic column density to variations in the ionizing radiation, respectively.
We obtained the values of $N_{i}$, $N_{i+1}$, $\alpha_{i-1}$, and $\alpha_{i-1}$ under varying ionization parameters in Cloudy simulations. Because these parameters vary with the depth of the cloud, we use the depth of each layer as a weight to obtain the final equivalent parameters. Panels d, e, and f of Fig. \ref{fig1} display the computed $|t_i^*|$ for $\rm C^{2+}$, $\rm C^{3+}$, and $\rm C^{4+}$, respectively. Interestingly, we discovered a discrepancy between the positive and negative response regions predicted by the $|t_i^*|$ and those determined through the peak of ion column density in the photoionization simulations, i.e., panel a versus d, b versus e, c versus f.

In this study, we conducted an in-depth examination of the physical mechanisms underpinning this misalignment and, in the process, identified a novel third response characteristic of ions, termed the hybrid response phenomenon, distinct from the conventional positive and negative responses. Section \ref{sec:simulation} presents time-dependent photoionization simulations and theoretical analysis. Section \ref{sec:conclusion} provides the conclusions.

\begin{figure}[htb]
\centering
\includegraphics[width=0.45\textwidth]{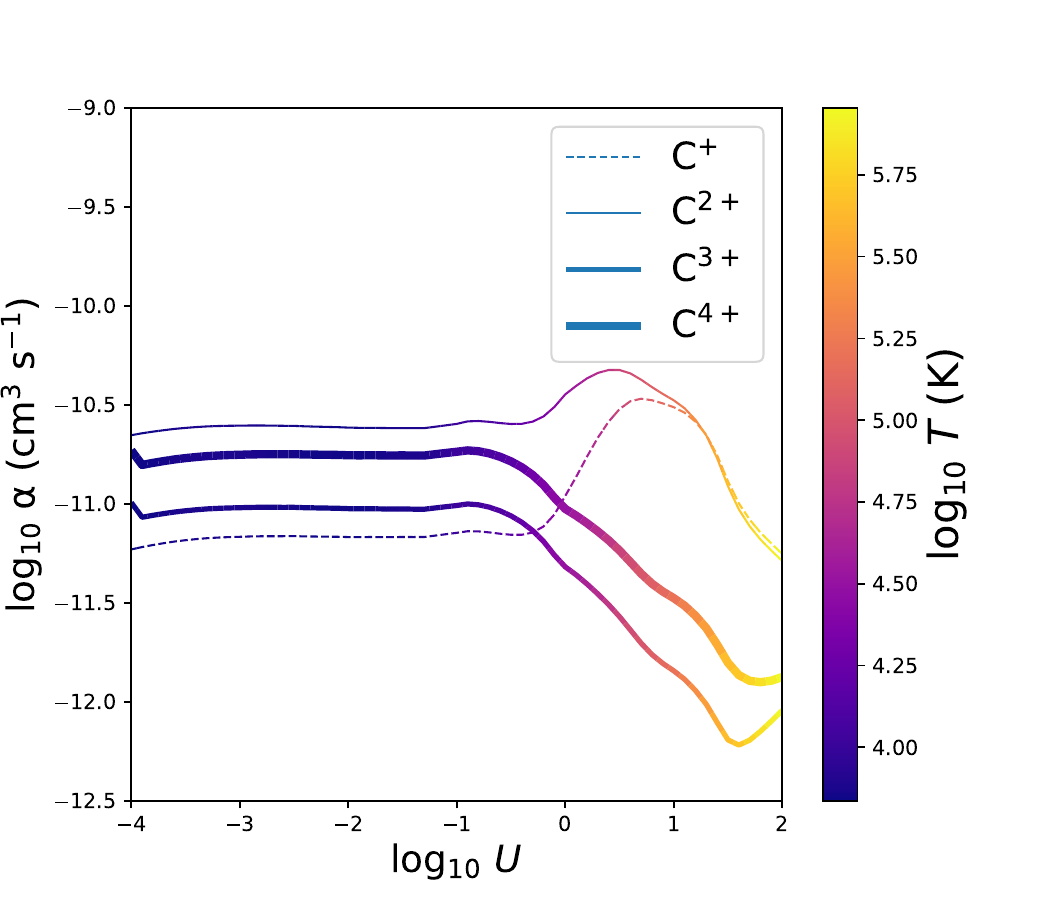}
\caption{\textbf{The Curve of Recombination Coefficient for Carbon Ions.} The curve, derived from photoionization simulations, is shown as a function of ionization parameters. The color bar represents the corresponding gas temperatures.
\label{fig2}}
\end{figure}

\begin{figure}[htb]
\centering
\includegraphics[width=0.45\textwidth]{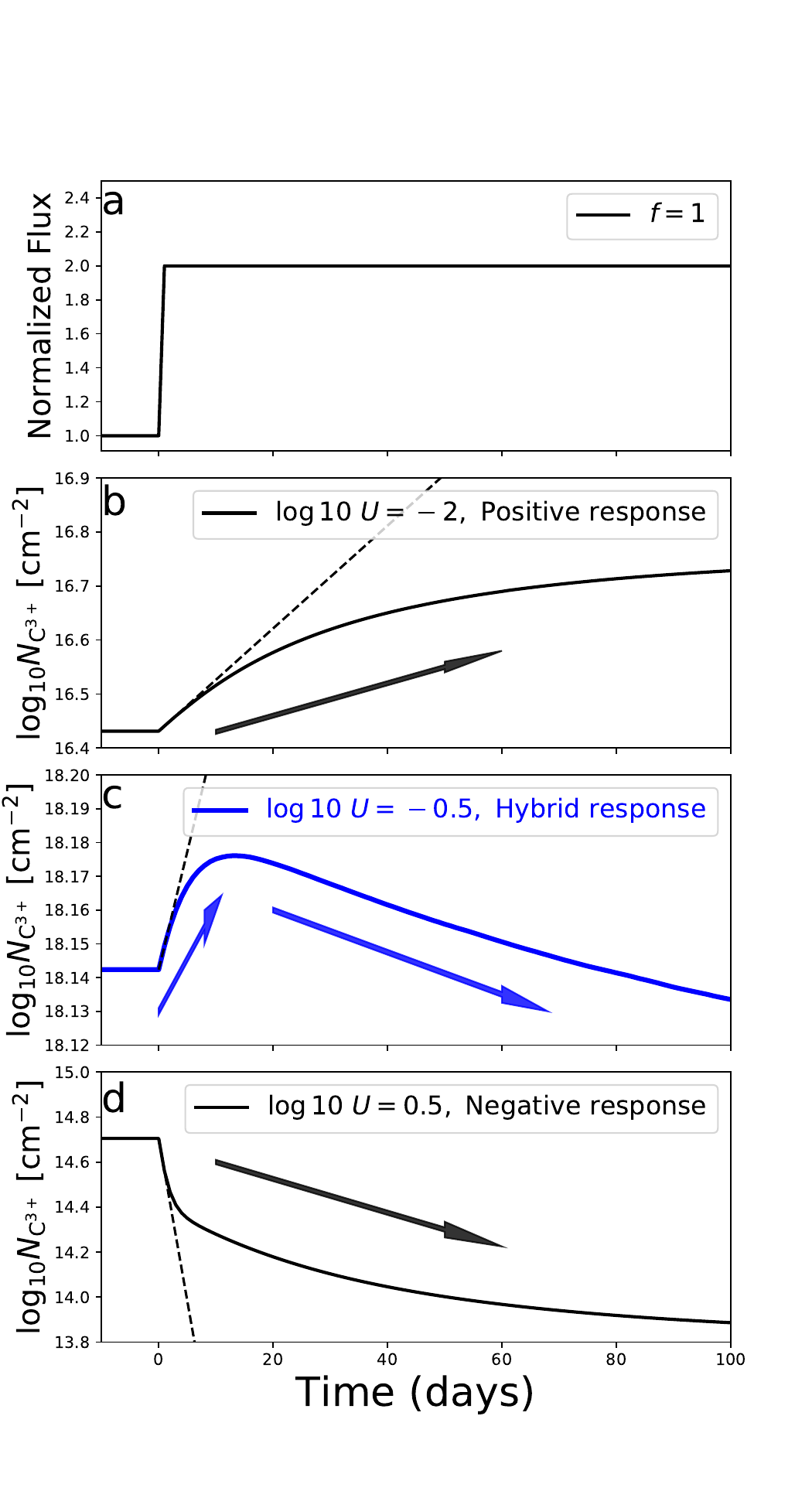}
\caption{\textbf{The Hybrid Response Effect for Carbon $\rm C^{3+}$ in Time-Dependent Photoionization Simulations.}
Panel (a) illustrates the sudden increase in ionizing radiation from 1.0 to 2.0 at $t=0$. Panel (b) shows a pure positive response at $\log_{10} \ U = -2.0$. Panel (c) demonstrates an apparent contradiction: the peak of column density predicts a negative response at $\log_{10} \ U = -0.5$, while the peak of the characteristic timescale predicts a positive response. Time-dependent photoionization simulations resolve this by revealing a hybrid response—initially exhibiting a temporary positive response before eventually transitioning to a negative response. Panel (d) displays a pure negative response at $\log_{10} \ U = 0.5$. The dashed lines represent the initial rate of change in column density as predicted by Equation \ref{eq5}.
\label{fig3}}
\end{figure}

\begin{figure}[htb]
\centering
\includegraphics[width=0.45\textwidth]{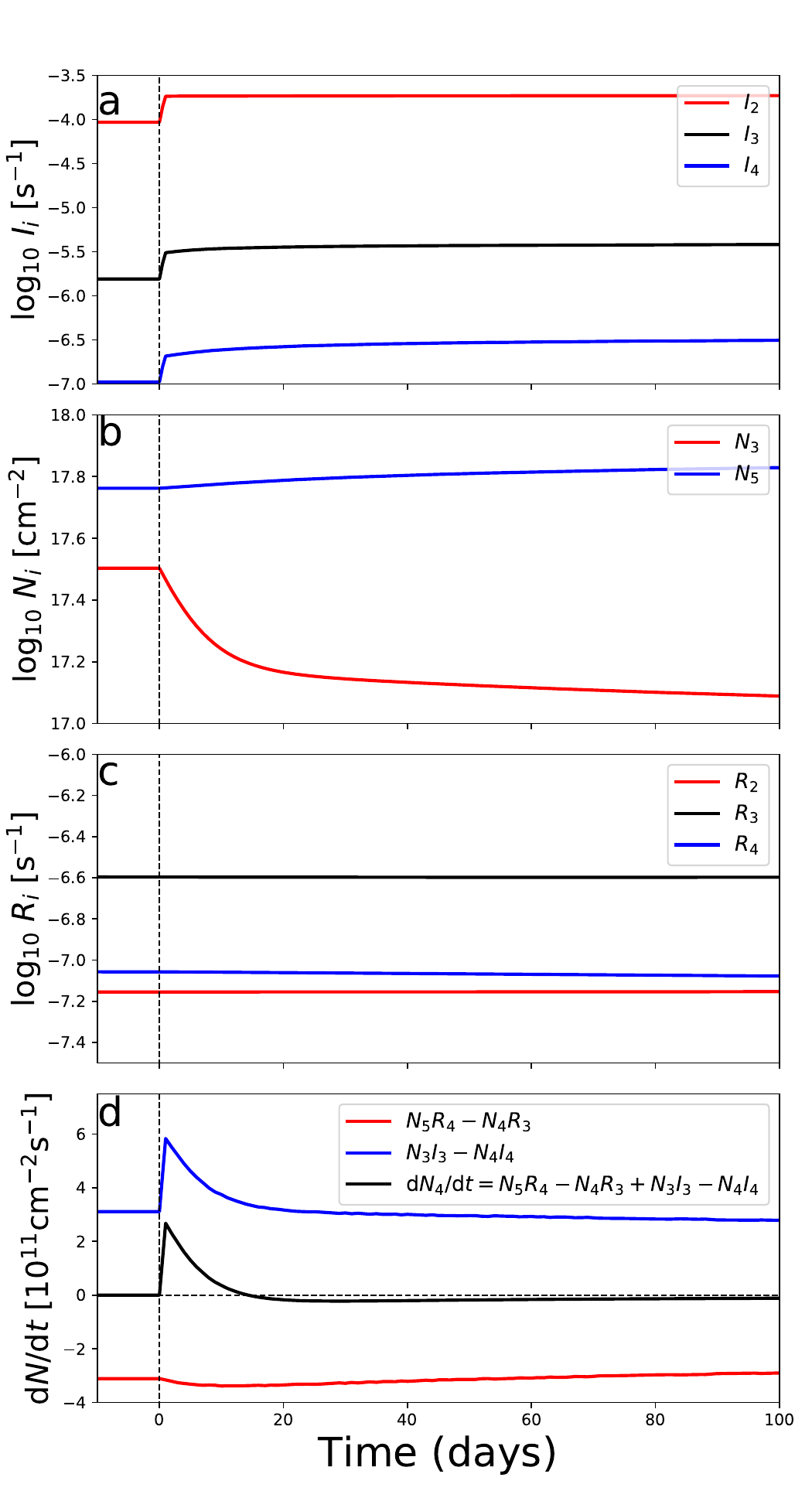}
\caption{\textbf{The evolution of ion column density, ionization rate, and recombination rate over time after a sudden change in radiation.}
Panel (a): after a sudden increase in radiation at t = 0, the column densities of ions do not exhibit a sudden change, but instead evolve gradually. Here $N_3$ refers to \ciii, and $N_5$ to \cv.  
Panel (b): the ionization rates undergo an abrupt change at t = 0.  
Panel (c): the recombination rates remain nearly constant throughout.  
Panel (d): time evolution of the rate of change in ion column density. In equilibrium, $N_3 I_3$ - $N_4 I_4$ is positive, and it also increases abruptly after the radiation rise, 
indicating that the initial rate of change in ion column density is positive.
\label{fig4}}
\end{figure}

\section{Photoionization simulations and theoretical analysis}
\label{sec:simulation}
In this section, we will begin with outlining the detailed parameter settings taken in our photoionization simulations. Then, we use the ion $\rm C^{3+}$ as a case study to illustrate the problem of positive and 
negative response misalignment through time-dependent photoionization simulations.

\subsection{Parameter settings in equilibrium state photoionization simulation}\label{sec:parameter}
We adopt typical parameters for broad absorption line (BAL) gas in quasars, with a gas density of $n_{\rm H} = 10^4\ \cc$ and a column density of $N_{\rm H} = 10^{22}\ \rm cm^{-2}$ \citep{he2019}. Using CLOUDY simulations \citep{ferland2017}, we compute models across a wide ionization parameter range, $-4 < \log_{10} \ U < 2$, with a step size of $\Delta \log_{10} \ U = 0.1$. The spectral energy distribution (SED) of the ionizing radiation source is UV-SOFT \citep{dunn2010a}, representative of high-luminosity, radio-quiet quasars.

Panels a, b, and c of Fig. \ref{fig1} show the column density curves for three ions ($\rm C^{2+}$, $\rm C^{3+}$, and $\rm C^{4+}$) as functions of the ionization parameter. The recombination coefficients, $\alpha_i$, for the carbon ions derived from the simulations are presented in Fig. \ref{fig2}. By substituting the simulated ion column densities and recombination coefficients into Equation \ref{eq3}, and assuming $f = 1.0$, we calculate the characteristic timescale (see bottom panels of Fig. \ref{fig1}).

The boundaries between positive and negative responses are indicated by vertical dashed lines in Fig. \ref{fig1}. Regardless of the ion species, it is evident that there is a discrepancy between the response regions predicted by Equation \ref{eq3} and those determined by the peaks of the ion column density in the photoionization simulations.

\subsection{Time-dependent photoionization simulations and the hybrid response phenomenon}

We use the trivalent ion of carbon, $\rm C^{3+}$, as an example to investigate the misalignment between positive and negative response boundaries through photoionization simulations. As shown in panels b and e of Fig. \ref{fig1}, the boundary determined by the peak of the column density is $\log_{10} \ U = -0.6$, while the boundary based on the characteristic timescale is $\log_{10} \ U = -0.05$. To examine the response behavior across these regions, we employ time-dependent CLOUDY simulations \citep{ferland2017} and focus on three intervals: $\log_{10} \ U < -0.6$, $-0.6 < \log_{10} \ U < -0.05$, and $\log_{10} \ U > -0.05$.

For this analysis, we select three specific ionization parameter values from these intervals: $\log_{10}\ U = -2$, $-0.5$, and $0.5$. According to panels b and e of Fig. \ref{fig1}, the column density of $\rm C^{3+}$ at $\log_{10}\ U = -2$ and $0.5$ is expected to exhibit positive and negative responses, respectively. However, for $\log_{10}\ U = -0.5$, the response behavior is unclear due to the apparent contradiction between the predictions in panels b and e of Fig. \ref{fig1}. To resolve this ambiguity, we perform a detailed time-dependent photoionization simulation (see the following paragraphs).

As shown in panel a of Fig.~\ref{fig3}, the initial incident ionizing flux is set to 1. At $t = 0$, the ionizing flux abruptly increases to 2.0 and remains constant thereafter, corresponding to an amplitude increase of $f = 1.0$. The time-dependent simulation spans a range of 0--100 days with a step size of 1 day.

Panels b and d of Fig.~\ref{fig3} illustrate the column density responses in the low-ionization region ($\log_{10} U = -2$) and high-ionization region ($\log_{10} U = 0.5$), respectively. In these cases, the column density exhibits pure positive responses in the low-ionization region and pure negative responses in the high-ionization region, consistent with expectations. However, as shown in panel c of Fig.~\ref{fig3}, at $\log_{10} U = -0.5$, the column density displays a hybrid response: an initial, temporary positive response followed by a transition to a negative response over time.

This finding reveals the presence of a third response mode---the hybrid response mode---alongside the purely positive and purely negative responses. The regions can thus be categorized based on ionization parameters as follows: pure positive response region (low ionization state), hybrid response region, and pure negative response region (high ionization state).

The existence of hybrid response regions significantly complicates the response process, making the relationship between observed radiation flux and absorption lines more intricate. For example, as depicted in the hybrid response of panel c in Fig.~\ref{fig3}, if the spectral signal-to-noise ratio and observation frequency are low, one might mistakenly interpret the column density as showing little to no response within the first 100 days following the increase in radiation, only noticing a significant decrease beyond that period.

\subsection{Explanation for hybrid response phenomenon}
Here, we analyze the physical mechanisms behind the existence of hybrid response regions. Firstly, in the ionization equilibrium state, the column density of ions is the result of the competition between ionization rate and recombination rate. The column density $N_{i}$, 
is a function of the column densities of adjacent valence ions, the corresponding ionization rates and recombination rates, expressed as $N_{i}=N_i(N_{i-1},N_{i+1},I_{i-1},I_{i},R_{i-1},R_{i})$. 
Assuming that the incident photon flux suddenly changes by a factor of 
$f$, only the ionization rate will experience an immediate corresponding change at that moment, 
i.e., $\d I_i$ = $fI_i$ and $\d I_{i-1}$ = $fI_{i-1}$ (Fig. \ref{fig4}a).
At this moment, the column densities of adjacent valence ions and recombination coefficients have not yet had the opportunity to adjust in response to the sudden change of the incident photon flux, i.e., $\d N_i$ = $\d N_{i-1}$ = $\d N_{i+1}$ = $\d R_i$ = $\d R_{i-1}$ = 0 (Fig. \ref{fig4}b, c). In particular, when the amplitude of the radiation change is small, the recombination rates remain nearly constants. Therefore, based on Equation \ref{eq1}, the evolution of $N_{i}$ can be written as:
\begin{eqnarray}  \label{eq4}
\frac{\d N_i}{\d t}=f(-N_{i}I_{i}+N_{i-1}I_{i-1}).
\end{eqnarray}
By applying Equation \ref{eq2}, one can further derive the rate of change of column density in logarithmic space at initial time:
\begin{eqnarray}  \label{eq5}
\frac{\d \ln N_i}{\d t}=\frac{f}{R_{i}}\left(-\frac{N_{i+1}}{N_{i}}+\frac{R_{i-1}}{R_{i}}\right).
\end{eqnarray}
From the above Equation, we can obtain the characteristic timescale of Equation \ref{eq3}.
Therefore, Equation \ref{eq4} governs the rate of change in column density at the initial moment, and $-N_{i}I_{i}+N_{i-1}I_{i-1}$ determines whether the response will be positive or negative during the early phase.  At the peak of ion concentration, the value of $-N_{i}I_{i}+N_{i-1}I_{i-1}$ is generally not equal to 0, and whether it is positive or negative varies for different ions. For example, for \civ, it is positive, but for \ciii, it is negative (see Fig. \ref{fig1}). 
The initial rate of change in column density predicted by Equation \ref{eq5} is represented by dashed lines in Fig. \ref{fig3}.
Note that Equations \ref{eq4} and \ref{eq5} apply only at the initial moment; they are no longer valid thereafter.
After enough time has passed, the gas will settle into a new equilibrium state, with both the recombination coefficient and ion column density stabilizing at new, determined values. The boundary of positive and negative responses in equilibrium state, that is, the position of peak column density, is determined by the recombination coefficient and ionization parameters in equilibrium state.
Therefore, at $\log_{10}\ U$ = -0.5, the early-stage positive response of $\rm C^{3+}$ aligns with the predictions based on the characteristic timescale (panel e of Fig. \ref{fig1}), while the subsequent negative response corresponds to the predictions of equilibrium-state photoionization simulations (panel b of Fig. \ref{fig1}). These two predictions are not contradictory; they just apply to different time ranges. The physical mechanism behind the hybrid response phenomenon is that: among the ionization rate, ion column density, and recombination rate, only the ionization rate can respond abruptly to changes in radiation. As a result, the initial rate of change in the column density of $N_i$  ion is determined by $-N_{i}I_{i}+N_{i-1}I_{i-1}$. However, at the peak of the $N_i$ ion column density, $-N_{i}I_{i}+N_{i-1}I_{i-1}$ is generally nonzero, leading to a misalignment between the boundaries of positive and negative responses. 

\section{Conclusion} 
\label{sec:conclusion}

Through photoionization simulations, we uncovered a puzzling phenomenon: the boundaries for positive and negative responses, as defined by the characteristic timescale, do not align with those determined by the peak ion column density. Further time-dependent photoionization simulations for carbon $\rm C^{3+}$ revealed that this discrepancy leads to the emergence of a third mode—the hybrid response mode—in addition to the pure positive and pure negative response modes.

Our analysis reveals that this phenomenon originates from the intrinsic asynchrony among the ionization rate, recombination rate, and ion column density. Among these, only the ionization rate can respond instantaneously to radiation changes. Consequently, the initial rate of change in the column density of \( N_i \) ions is given by \( -N_i I_i + N_{i-1} I_{i-1} \). However, this quantity is typically nonzero at the peak of \( N_i \), resulting in a misalignment between the transition boundaries of positive and negative responses.

The hybrid response introduces complexity into interpreting ion column density responses to ionizing radiation in observational studies, making it more challenging to measure or constrain physical properties such as gas density and ionization parameters. However, this phenomenon might also provide novel opportunities for probing gas properties.

\acknowledgments
Zhicheng He is supported by the National Natural Science Foundation of China (nos. 12222304, 12192220, and 12192221).


\end{document}